\input epsf
\documentstyle[pre,aps]{revtex}
\begin{document}

\twocolumn[
\hsize\textwidth\columnwidth\hsize\csname @twocolumnfalse\endcsname

\date{\today}
\title{Non-equilibrium Phase-Ordering with a Global Conservation Law}
\author{A. D. Rutenberg \cite{email}}
\address{Theoretical Physics, University of Oxford, Oxford OX1 3NP, United
Kingdom}
\maketitle

\widetext

\begin{abstract}
In all dimensions, infinite-range Kawasaki spin exchange in a 
quenched Ising model leads to an asymptotic length-scale
$L \sim (\rho t)^{1/2} \sim t^{1/3}$ at $T=0$ because the kinetic 
coefficient is 
renormalized by the broken-bond density, $\rho \sim L^{-1}$.
For $T>0$, activated kinetics recovers
the standard asymptotic growth-law, $L \sim t^{1/2}$. However, at all
temperatures, infinite-range energy-transport is allowed by the 
spin-exchange dynamics.
A better implementation of global conservation, 
the microcanonical Creutz algorithm, is well behaved and
exhibits the standard non-conserved growth law, $L \sim t^{1/2}$, 
at all temperatures.
\end{abstract}
\narrowtext
]

Globally conserved dynamics offers a useful test of our theoretical 
understanding of phase-ordering \cite{Bray94}
because it allows access to off-critical quenches without requiring
diffusive ``model-B'' transport of the order-parameter. 
RG \cite{Bray89}
arguments show that adding global conservation changes neither the 
critical dynamics nor the asymptotic growth law after a quench 
into the ordered phase. This has been confirmed by 
numerical studies \cite{Annett92}. The only change to non-conserved 
dynamics is to impose $\partial_t \langle \phi \rangle=0$ where, using spin
language, $\phi$ is the local magnetization ($\langle \phi \rangle$ is
its spatial average). The standard ``model-A'' dissipative dynamics are
$\partial_t \phi = - \Gamma \delta F/ \delta \phi$, where 
$F[\{\phi\}]= \int d^d r [ (\nabla \phi)^2 + (\phi^2-1)^2 ]$. The 
global constraint is imposed with a 
a uniform magnetic field term, $H \langle \phi \rangle$, added
to $F$.  The field $H(t)$, effectively a lagrange multiplier, is
decreased smoothly while the ordering progresses so as to 
maintain $\langle \phi \rangle$ constant. 

With a global conservation law, $\langle \phi \rangle$ is a 
tunable parameter that does not change the growth-law but does affect
scaled correlation functions and other aspects of the system. One can
use this to explore what ``universality'' entails in non-equilibrium
phase-ordering systems.  
For example, with global conservation in Ising models, 
Sire and Majumdar \cite{Sire95} have shown that the autocorrelation
exponent $\lambda$ depends on the net magnetization.
This author \cite{Rutenberg96} has found spatial anisotropy 
in scalar phase-ordering systems that depends strongly on 
$\langle \phi \rangle$.

However, there has been uncertainty about how to implement 
global conservation.  Tuning a magnetic field to maintain 
$\langle \phi \rangle$ is 
impractical in computer simulations, due to the stochastic
nature of most algorithms. Two alternative algorithms
have been used for Ising models: 
a Creutz demon \cite{Creutz83}, and infinite-range
Kawasaki spin-exchange \cite{Annett92,Sire95,Tamayo89}.  
Differences between these
two implementations at low temperatures has led to some confusion in 
the past \cite{Sire95,Tamayo89,Goryachev94}. In this report 
we clarify the situation. 

For a 2d Ising model on a square lattice, 
consider infinite-range Kawasaki exchange dynamics
\cite{Annett92,Sire95,Tamayo89}: two randomly
selected spins are exchanged under heat-bath dynamics, i.e. with probability
$1/[1+\exp(\Delta E/k_BT)]$, where $\Delta E$ is the energy change under
the spin exchange.  The exchange satisfies global conservation.
However at low temperature the asymptotic dynamics are {\em activated},
and cross-over to a different growth law at zero-temperature. 

At $T=0$, no exchanges that increase the 
energy are accepted.   Starting from random initial conditions, 
aggregation of spins takes place and
the broken-bond density $\rho(t)$ will be proportional to the domain
wall density at late times, $\rho \sim L^{-1}$. 
Isolated spins with all bonds broken, of number-density $N$, 
will aggregate rapidly 
onto existing domains, $\partial_t N \propto - \rho N$, and may 
be ignored.  Spins which are part of domain walls
can exchange, but only with the fraction 
$\rho$ of the system that has broken bonds (i.e. that are also
on domain walls).  This rescales the effective kinetic coefficient
by a factor of $\rho \sim  L^{-1}$. 
For $T=0$ infinite-range Kawasaki exchange we thus 
expect $L \sim (\rho t)^{1/2} \sim (t/L)^{1/2} \sim t^{1/3}$, 
where $t^{1/2}$ is the
standard non-conserved growth \cite{Bray94}.
This applies in any dimension.

For $T>0$, thermal fluctuations provide partners for 
spin exchange and also allow exchange with bulk spins. 
These activated processes 
will dominate after the broken-bond density 
becomes comparable to the equilibrium average.
After a quench, this will result in a crossover from intermediate 
$L\sim t^{1/3}$ growth to asymptotic $L \sim t^{1/2}$ growth.
[The length-scale is extracted from
the scaling of the spherically-averaged correlation function
$C(r,t) = \langle \phi(x) \phi(x+r) \rangle = M^2 f(r/L)$, where $M$ is
the equilibrium bulk magnetization. $L$ is chosen so that $f(1)=1/2$.] 
Results for quenches to $T_c$ will be similarly affected
--- the dynamical exponent will be underestimated while
$\rho$ relaxes towards equilibrium \cite{perhaps}.

A renormalized kinetic coefficient alone will not affect {\em scaled}
correlations. However Kawasaki exchange
also allows infinite-range energy-transfer through the system.  
For example, at $T=0$, an isolated spin (four broken bonds) 
may hop freely through the system by spin
exchange with bulk spins (no broken bonds). Similarly,
a ``bump'' on a flat interface (with one broken bond) may hop to any
other flat interface of the system. Effects of this energy transport
might show up 
at low-temperatures in non-universal effects such as spatial
anisotropy in the late-stages of the quench \cite{Rutenberg96}. It is
unclear if this would affect the autocorrelation results of Sire
and Majumdar \cite{Sire95}, who used this algorithm at $T=0$. 

A faster and safer algorithm was proposed by Creutz \cite{Creutz83}.
The system is coupled to a small spin reservoir (here of size 2).
Single spin flips accepted by normal kinetic
Ising model dynamics are subject to the additional condition that the 
required spin change ($\pm 2$) can be extracted from the reservoir.
The dynamics of the system plus reservoir are microcanonical in the 
magnetization, and there are none of the questions of 
energy-transport that we have just raised for infinite-range
Kawasaki exchange \cite{warning}. As a result the RG
results directly apply and a growth-law of $L \sim t^{1/2}$ is
indeed seen \cite{Annett92}.

\begin{figure}
\begin{center}
\mbox{
\epsfxsize=3.5in
\epsfbox{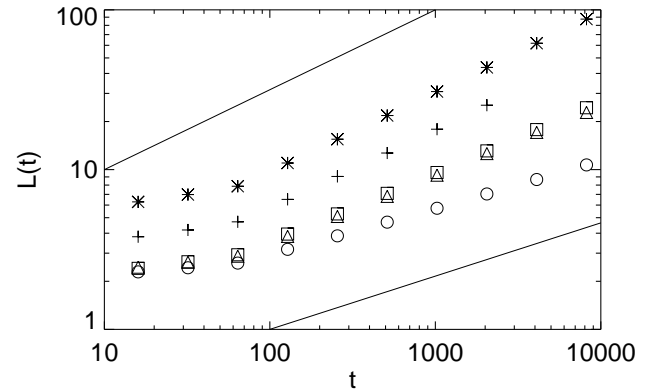}}
\end{center}
\caption{Length-scale $L(t)$ for non-conserved
(stars), Creutz (plus), Kawasaki $T=0.9 T_c$ (triangle), and Kawasaki
$T=0$ (circle).  
The system sizes are $2048^2$, $1024^2$, $512^2$,
and $512^2$, respectively, with at least $20$ samples in each system.
We also show $L / \protect\sqrt{\rho}$
for the $T=0$ Kawasaki quench (squares), which goes as $t^{1/2}$, as expected. 
The upper and lower lines show $t^{1/2}$ and $t^{1/3}$, respectively.
\label{FIG:lengths}}
\end{figure}

\begin{figure}
\begin{center}
\mbox{
\epsfxsize=3.5in
\epsfbox{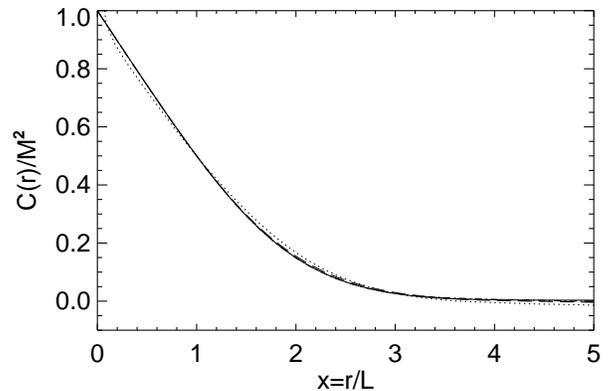}}
\end{center}
\caption{Scaled correlations, divided by the equilibrium magnetization
squared, from the latest times of 
simulations shown in the previous figure. Solid, dashed, dot-dashed, and
dotted lines are non-conserved, Creutz $T=0$, Kawasaki $T= 0$, and
Kawasaki $T=0.9 T_c$, respectively.  
\label{FIG:corr}}
\end{figure}

To check these arguments we simulated critical quenches, with 
$\langle \phi \rangle = 0$, 
for non-conserved and also for globally conserved systems with both
Creutz and Kawasaki dynamics. For off-critical
quenches similar results apply. In Fig. \ref{FIG:lengths} we see the 
expected $t^{1/3}$ growth for $T=0$ Kawasaki exchange, and reconfirm $t^{1/2}$
growth for the other cases.  To test the rescaling of the kinetic prefactor
in the Kawasaki algorithm, we show a quench to $T=0.9 T_c$ and also $L /
\sqrt{\rho}$ for the quench to $T=0$. Both of these exhibit the expected
$t^{1/2}$ growth. In Fig. \ref{FIG:corr}, we
see that all of the models have similar spherically-averaged scaled 
correlations. [Anisotropies are too small to 
see for these critical quenches \cite{Rutenberg96}.] The correlations for
the Kawasaki model are slightly different, however 
this seems to be a transient effect. [Strong finite-size effects in the
$T=0.9 T_c$ quench, after the latest time shown, limit the duration of
the simulation.]

In conclusion, we see that an infinite-range Kawasaki exchange
implementation of global conservation laws has
activated kinetics, with an anomalous $L \sim t^{1/3}$ growth 
at $T=0$. In addition, Kawasaki dynamics leads to long-range energy
transport. The Creutz algorithm, on the other hand, is not activated and has no
long-range energy transport. It provides
a faster and better-controlled implementation of global conservation
laws, and should exhibit the expected $t^{1/2}$ growth at all temperatures.
 
This work was supported by EPSRC grant GR/J78044.

\begin{references}
\bibitem{perhaps} This may be the cause of the anomalously small
 dynamical critical exponent seen by Tamayo and Klein 
\protect\cite{Tamayo89} with
infinite-range Kawasaki exchange.
\bibitem{warning} The Creutz algorithm must not be
implemented with a sublattice update since the sweep 
direction generates anomalous spatial anisotropies.
\end{references}
\end{document}